\documentclass[
  aps,
  prl,
  reprint,
  superscriptaddress,
  amsmath,
  amssymb
]{revtex4-2}

\usepackage{graphicx}
\usepackage{bm}
\usepackage{hyperref}
\usepackage{xcolor}

\begin{document}

\title{Origins of Faceted Anomalous Scaling}

\author{Maximilien Bernard}
\affiliation{Universit\'e Paris-Saclay, CNRS, LPTMS, 91405 Orsay, France}
\affiliation{Laboratoire de Physique de l'\'Ecole Normale Sup\'erieure, ENS, Universit\'e PSL, CNRS, Sorbonne Universit\'e, Universit\'e Paris Cit\'e, 75005 Paris, France}

\author{Jos\'e Luis Iguain}
\affiliation{Instituto de Investigaciones F\'isicas de Mar del Plata (IFIMAR), CONICET and Universidad Nacional de Mar del Plata, B7602AYL Mar del Plata, Argentina}

\author{Alejandro Kolton}
\affiliation{Centro At\'omico Bariloche and Instituto Balseiro, CNEA, CONICET and Universidad Nacional de Cuyo, R8402AGP San Carlos de Bariloche, Argentina}

\author{Alberto Rosso}
\affiliation{Universit\'e Paris-Saclay, CNRS, LPTMS, 91405 Orsay, France}

\date{\today}

\begin{abstract}
Fluctuating interfaces are typically characterized by their global, spectral, and local roughness exponents.
Faceted anomalous scaling, characterized by $\alpha_s>1$, $\alpha_s\neq\alpha$, and $\alpha_{\mathrm{loc}}=1$, owes its name to the tendency of such interfaces to develop long, nearly linear segments called facets.
Its statistical origin, and whether facets are required, have remained unclear.
We show that this exponent structure is determined by the spatial statistics of the slope field.
Slope increments split into typical fluctuations and rare, larger jumps.
This anomalous scaling arises either when the typical slope increment displays genuine system-size-dependent anomalous scaling or when the rare jumps dominate the second moment.
The two contributions can coexist, and their competition leads to bifractal scaling of the slopes.
We establish this framework analytically in the generalized anharmonic Larkin model and the columnar Kardar--Parisi--Zhang equation, and we derive exactly the global, spectral, and local roughness exponents in both cases.
The Larkin model realizes both mechanisms and produces anomalous interfaces with or without facets, whereas columnar KPZ belongs to the jump-dominated case and yields $\alpha_s=3/2$.
\end{abstract}

\maketitle

\begin{figure*}[t]
  \includegraphics[width=\textwidth]{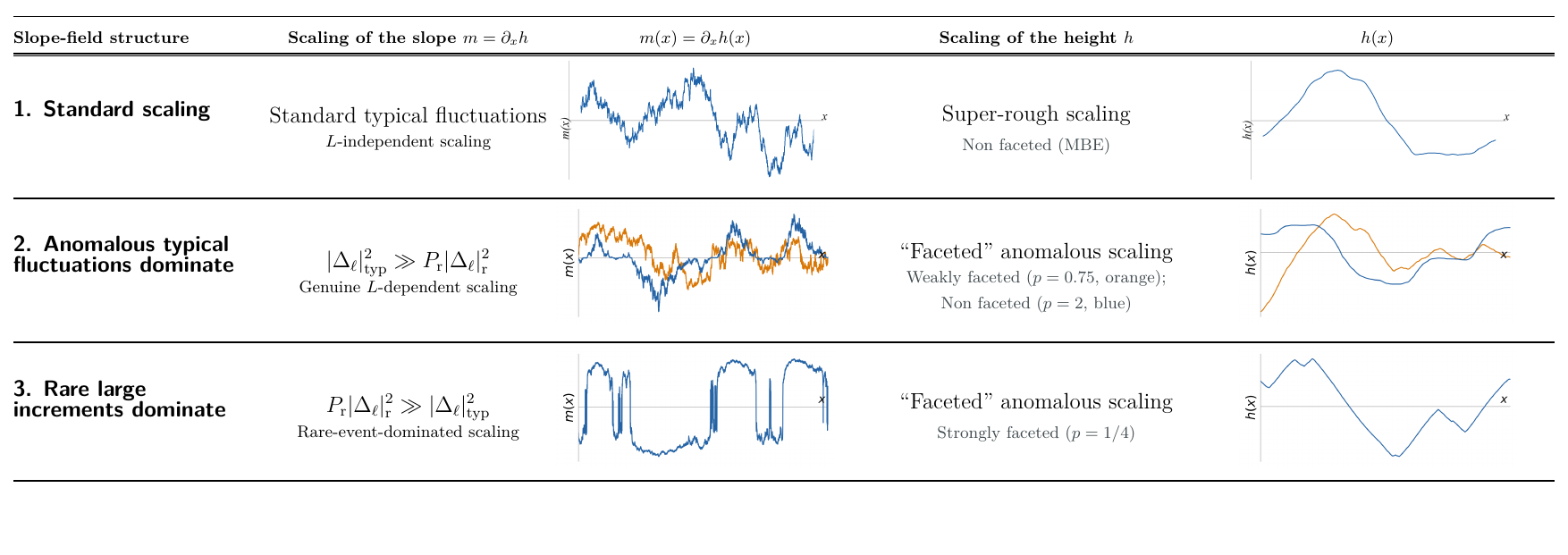}
  \caption{Phenomenological slope-field classification of stationary one-dimensional rough interfaces with $\alpha_{\mathrm{loc}}=1$ and $\alpha_s>1$, according to the contribution controlling the second moment of slope increments.
  Here $\Delta_\ell=m(x+\ell)-m(x)$, and Eq.~\eqref{eq:slope-sector-decomposition} decomposes $\langle|\Delta_\ell|^2\rangle$ into typical and rare contributions, the latter weighted by the probability $P_{\mathrm r}(\ell,L)$ of sampling a large rare increment.
  Standard $L$-independent typical fluctuations yield a super-rough, nonfaceted height field (top).
  ``Faceted'' anomalous scaling has two origins, illustrated by the anharmonic Larkin model: dominant anomalous typical fluctuations produce weakly faceted ($1/2<p<1$) or nonfaceted ($p>1$) morphologies (middle), whereas dominant rare large increments produce strong faceting ($0<p<1/2$; bottom).
  Columnar KPZ also realizes the rare-dominated class and is analyzed separately in Fig.~\ref{fig:columnar-kpz-structure-factor}.}
  \label{fig:slope-classification}
\end{figure*}

Scale invariance underlies the classification of fluctuating interfaces.
In the Family--Vicsek picture, a self-affine height field $h(x,t)$ is characterized by a single roughness exponent $\alpha$ and a growing correlation length $\xi(t)\sim t^{1/z}$~\cite{FamilyVicsek1985}.
This description includes paradigmatic models such as the Edwards--Wilkinson (EW)~\cite{EdwardsWilkinson1982} and Kardar--Parisi--Zhang (KPZ)~\cite{KardarParisiZhang1986} equations with space-time white noise.
We consider stationary interfaces, for which $\xi(t)\simeq L$ and the time argument can be omitted.
Even then, roughness may require three distinct exponents~\cite{LopezRodriguezCuerno1997Power}.
The global roughness exponent $\alpha$ is defined by
\begin{equation}
  W^2(L)
  \equiv
  \left\langle \frac{1}{L}\int_0^L dx\,[h(x)-\bar h]^2\right\rangle
  \sim L^{2\alpha},
  \label{eq:stationary-global-scaling}
\end{equation}
where $\bar h=L^{-1}\int_0^L dx\,h(x)$.

The local exponent $\alpha_{\mathrm{loc}}$ follows from height increments over $\ell\ll L$~\cite{LopezRodriguezCuerno1997Power,Lopez1999Scaling},
\begin{equation}
  G_2(\ell,L)
  \equiv
  \left\langle \left|h(x+\ell)-h(x)\right|^2 \right\rangle
  \sim
  \ell^{2\alpha_{\mathrm{loc}}}
  L^{2\alpha-2\alpha_{\mathrm{loc}}}.
  \label{eq:stationary-real-space-scaling}
\end{equation}
Finally, defining $\hat h(k)=L^{-1/2}\int_0^L dx\,e^{-ikx}[h(x)-\bar h]$, the spectral exponent $\alpha_s$ is given by
\begin{equation}
  S(k,L)
  \equiv
  \left\langle \left|\hat h(k)\right|^2 \right\rangle
  \sim
  k^{-(1+2\alpha_s)}
  L^{2(\alpha-\alpha_s)}
  \label{eq:stationary-spectral-scaling}
\end{equation}
for $1/L\ll k\ll1$.

Family--Vicsek scaling corresponds to $\alpha=\alpha_{\mathrm{loc}}=\alpha_s$.
Two well-understood departures from Family--Vicsek scaling are known.
In the first, $\alpha_s=\alpha_{\mathrm{loc}}<1$ while $\alpha\neq\alpha_s$~\cite{LopezRodriguezCuerno1997Power,Lopez1999Scaling,Auger2006TiN,Balankin2006Crumpled,LopezSchmittbuhl1998Fracture,Soriano2002HeleShaw,Bernard2024Heterogeneous,bernard2025stochastic,bernard2026fluctuations}.
In super-rough interfaces, such as the quenched Edwards--Wilkinson equation at the depinning threshold~\cite{PhysRevLett.70.2973,rosso2001origin} and linear molecular beam epitaxy~\cite{amar1993groove,sarma1994kinetic}, $\alpha_s=\alpha>1$ while $\alpha_{\mathrm{loc}}=1$.

A third departure, known as \emph{faceted anomalous scaling}, is characterized by $\alpha_s>1$, $\alpha_s\neq\alpha$, and $\alpha_{\mathrm{loc}}=1$.
First observed in the Sneppen model~\cite{Sneppen1992SelfOrganized}, this exponent structure has since been reported in growth models involving localization~\cite{Szendro2007Localization}, temporally correlated noise~\cite{Ales2019Faceted}, and nonlinear elasticity~\cite{Purrello2019Larkin,Ales2021Anharmonic}, as well as experimentally in dissolving metal surfaces and epitaxially grown semiconductor films~\cite{CordobaTorres2009Metal,Nascimento2011Faceted}.
Unlike the other two departures, however, its statistical origin remains unclear.
The term ``faceted'' refers to the frequent appearance of long, nearly linear segments, but morphology alone neither explains why $\alpha_s$ differs from both $\alpha$ and $\alpha_{\mathrm{loc}}$ nor establishes whether facets are required for this exponent structure.
The key observation is that this exponent structure is controlled by the spatial statistics of the slope field $m(x)\equiv\partial_x h(x)$: slope increments separate into typical fluctuations and rare, sharper events, and their competition governs all three roughness exponents.

In this Letter, we establish the statistical origin of this exponent structure by treating the stationary local slope, \(m(x)\equiv\partial_x h(x)\), as a fluctuating interface in its own right.
We show that the separation of \(\alpha_s\) from both \(\alpha\) and \(\alpha_{\mathrm{loc}}\) can originate from two contributions to slope-increment moments: typical increments with genuine system-size-dependent scaling, and rare large increments whose sampling probability may vanish as \(L\to\infty\).
When the latter dominate the second moment, the associated large slope changes generate a pronounced faceted morphology; when the anomalous typical contribution dominates, the same exponent structure does not require facets.
The two contributions may coexist and exchange dominance with moment order, naturally producing bifractal slope statistics.
Their relation to height morphology is summarized in Fig.~\ref{fig:slope-classification}.
One-dimensional columnar KPZ isolates the rare-increment-dominated limit and yields \(\alpha_s=3/2\).
The generalized anharmonic Larkin model instead tunes the competition between the two contributions and the degree of visible faceting.
For both models, we derive exact expressions for the global, spectral, and local roughness exponents.

\paragraph{Spectral constraints and slope scaling.}
Using Eq.~\eqref{eq:stationary-spectral-scaling} in the Fourier representation of the height increment gives
\begin{align}
  G_2(\ell,L)
  &\propto
  \int_{1/L}^{1}dk\,[1-\cos(k\ell)]S(k,L)
  \nonumber\\
  &\sim
  \begin{cases}
    \ell^{2\alpha_s}L^{2(\alpha-\alpha_s)}, & \alpha_s<1,\\
    \ell^2L^{2(\alpha-1)}, & \alpha_s>1.
  \end{cases}
  \label{eq:local-exponent-from-spectrum}
\end{align}
Hence $\alpha_{\mathrm{loc}}=\alpha_s$ for $\alpha_s<1$, whereas $\alpha_{\mathrm{loc}}=1$ for $\alpha_s>1$, with logarithmic corrections at $\alpha_s=1$.
Let $\alpha_m$, $\alpha_{s,m}$, and $\alpha_{\mathrm{loc},m}$ denote the corresponding slope-field exponents.
The slope-field analogue of Eq.~\eqref{eq:stationary-real-space-scaling} then reads
\begin{equation}
  G_2^m(\ell,L)
  \equiv
  \left\langle|\Delta_\ell|^2\right\rangle
  \sim
  \ell^{2\alpha_{s,m}}L^{2(\alpha_m-\alpha_{s,m})},
  \label{eq:slope-scaling}
\end{equation}
where $\Delta_\ell=m(x+\ell)-m(x)$; the slope exponents are identified by matching this scaling. For periodic boundaries, \(\hat m_k=ik\hat h_k\). For free boundaries, the corresponding cosine modes of \(h\) and sine modes of \(m\) satisfy \(m_n^{(s)}=-k_nh_n^{(c)}\). In either case, Eq. (4) applied to both fields gives
\begin{equation}
  \begin{aligned}
    \alpha &= 1+\alpha_m,\\
    \alpha_s &= 1+\alpha_{s,m},\\
    \alpha_{\mathrm{loc}} &= 1
    \quad\bigl(0<\alpha_{\mathrm{loc},m}=\alpha_{s,m}<1\bigr),
  \end{aligned}
  \label{eq:height-slope-exponent-relations}
\end{equation}
where the last relation applies to the range $1<\alpha_s<2$ considered below.
Both super-rough and faceted anomalous interfaces obey these relations.
For super-rough interfaces, standard scaling of typical slope fluctuations implies $\alpha_m=\alpha_{s,m}=\alpha_{\mathrm{loc},m}>0$, and hence $\alpha=\alpha_s>1$~\cite{Lopez1999Scaling,LopezCastroGallego2005Slopes,Ales2020Superroughening}.
Faceted anomalous scaling instead requires $\alpha_m\neq\alpha_{s,m}$, and therefore nonstandard slope statistics.
To identify their possible origins, define $\Delta_\ell=m(x+\ell)-m(x)$ and decompose its second moment, at the scaling level, as
\begin{equation}
  \left\langle|\Delta_\ell|^2\right\rangle
  \sim
  |\Delta_\ell|_{\mathrm{typ}}^2
  + P_{\mathrm r}(\ell,L)|\Delta_\ell|_{\mathrm r}^2 .
  \label{eq:slope-sector-decomposition}
\end{equation}
The first term represents typical slope increments; the second represents rare large increments of characteristic amplitude $|\Delta_\ell|_{\mathrm r}$, sampled with a probability $P_{\mathrm r}(\ell,L)$ that may vanish as $L\to\infty$.
Faceted anomalous scaling can thus arise either from an $L$-dependent typical contribution or from dominance of the rare contribution.
We isolate the latter mechanism in columnar KPZ before studying their competition in the generalized Larkin model.

\paragraph{Scaling dominated by rare large increments: columnar KPZ.}
We illustrate this limit with the one-dimensional columnar KPZ equation~\cite{Szendro2007Localization},
\begin{equation}
  \partial_t h(x,t)
  =\partial_x^2h(x,t)
  +\lambda[\partial_xh(x,t)]^2
  +\eta(x),
  \label{eq:columnar-kpz}
\end{equation}
whose defining feature is that the random force \(\eta(x)\) is independent of time.
Under the Cole--Hopf transformation \(Z=e^{\lambda h}\), Eq.~\eqref{eq:columnar-kpz} becomes the imaginary-time Schr\"odinger equation
\begin{equation}
  \partial_t Z(x,t)=-\mathcal H Z(x,t),
  \qquad
  \mathcal H=-\partial_x^2-\lambda\eta(x),
  \label{eq:columnar-kpz-cole-hopf}
\end{equation}
where the time-independent random operator \(\mathcal H\) is the one-dimensional Anderson Hamiltonian.
Let \(\mathcal H\psi_n=E_n\psi_n\), with \(E_0\) the ground-state energy.
The spectral decomposition then gives
\begin{equation}
  Z(x,t)=\sum_n c_n\psi_n(x)e^{-E_nt}
  \underset{t\to\infty}{\simeq}
  c_0\psi_0(x)e^{-E_0t},
  \label{eq:columnar-kpz-spectral-decomposition}
\end{equation}
so that the interface reaches a stationary profile in a frame moving at velocity \(v_\infty=-E_0/\lambda\):
\begin{equation}
  H(x)\equiv h(x,t)-v_\infty t
  \simeq \frac{1}{\lambda}\ln\psi_0(x)+\mathrm{const}.
  \label{eq:columnar-kpz-stationary-profile}
\end{equation}
In one dimension, the ground state is exponentially localized around a random position \(x_c\)~\cite{NattermannRenz1989RandomEnvironment,KrugHalpinHealy1993Columnar,Texier2000EnergyLevels,DumazLabbe2020Localization}.
Its logarithm can therefore be decomposed as
\begin{equation}
  \ln\psi_0(x)=-\gamma|x-x_c|+\chi(x),
  \label{eq:columnar-kpz-localized-state}
\end{equation}
where \(\gamma\) is the inverse localization length and \(\chi\) describes subleading spatial fluctuations.
The envelope produces two long, nearly linear segments joined near \(x_c\).
Their vertical span is of order \(\gamma L\); hence its leading algebraic dependence is \(W(L)\sim L\), giving \(\alpha=1\).

\begin{figure}[t]
  \includegraphics[width=\columnwidth]{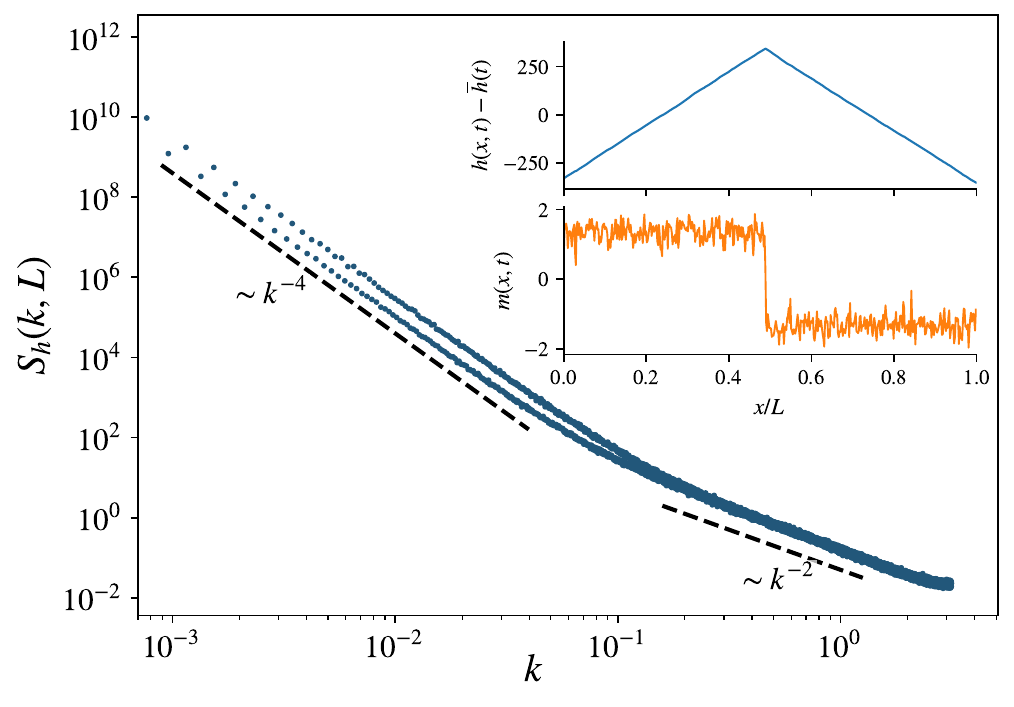}
  \caption{Stationary height structure factor \(S_h(k,L)\) of the one-dimensional columnar KPZ equation for \(L=2^{14}\), averaged over 360 disorder realizations, with free boundary conditions.
  The dashed segments indicate the \(k^{-4}\) behavior at small \(k\), generated by the macroscopic slope jump, and the \(k^{-2}\) behavior at larger \(k\), associated with the short-scale fluctuations \(\chi(x)\) of the localized state.
  Insets: centered height profile \(h(x,t)-\overline{h}(t)\) (upper) and corresponding slope profile \(m(x,t)=\partial_x h(x,t)\) (lower) for a representative stationary configuration.
  The two finite slope plateaus separated by a macroscopic jump produce the faceted height profile.
  The structure factor is computed with the cosine transform appropriate to free boundary conditions, and its oscillatory sub-branches, which arise from profiles with one or a few kinks, do not affect the \(k^{-4}\) envelope.}
  \label{fig:columnar-kpz-structure-factor}
\end{figure}

Consider now the stationary slope for free boundaries,
\(m(x)=\partial_xH(x)=\lambda^{-1}\partial_x\ln\psi_0(x)\), which has two nearly constant plateaus separated by a single jump at \(x_c\).
Denoting its amplitude by \(J\equiv m(x_c^+)-m(x_c^-)\), an interval of length \(\ell\ll L\) crosses the jump with probability \(P_{\mathrm r}\sim\ell/L\).
The contribution from rare large increments controls the large-scale behavior and gives
\begin{equation}
  \left\langle|\Delta_\ell|^2\right\rangle
  \sim \left\langle J^2\right\rangle\frac{\ell}{L}.
  \label{eq:jump-slope-increments}
\end{equation}
The \(\ell\) and \(L\) dependences in Eq.~\eqref{eq:jump-slope-increments} give \(\alpha_{\mathrm{loc},m}=1/2\) and \(\alpha_m=0\).
Equation~\eqref{eq:height-slope-exponent-relations} then yields \(\alpha_{s,m}=1/2\) and the complete height-exponent triplet
\begin{equation}
  (\alpha,\alpha_s,\alpha_{\mathrm{loc}})
  =\left(1,\frac32,1\right).
  \label{eq:columnar-kpz-exponents}
\end{equation}
The predicted small-$k$ decay $S_h(k,L)\sim k^{-4}$, generated by the macroscopic slope jump, is confirmed in Fig.~\ref{fig:columnar-kpz-structure-factor}.
The $k^{-2}$ regime at larger $k$, associated with the Brownian-like fluctuations \(\chi(x)\), is discussed in the Supplemental Material.
The argument extends beyond columnar KPZ whenever a finite number of macroscopic slope jumps dominates the second moment, including in heterogeneous elastic lines~\cite{Bernard2024Heterogeneous}.
As detailed in the End Matter, the same result extends to any dimension under the geometric assumption of a finite number of macroscopic planar facets.

We now turn to the anharmonic Larkin model, where the two terms in Eq.~\eqref{eq:slope-sector-decomposition} can coexist and compete.

\paragraph{Competing slope fluctuations: anharmonic Larkin model.}
We consider the one-dimensional anharmonic Larkin model~\cite{Purrello2019Larkin},
\begin{equation}
  \partial_t h(x,t)
  = \partial_x\!\left[(\partial_xh)^{2n-1}\right]
  +\eta(x),
  \qquad n=2,3,\ldots,
  \label{eq:alm}
\end{equation}
where \(\eta(x)\) is a quenched spatial white noise.
For a stationary translating solution \(h(x,t)=vt+H(x)\), with \(m(x)=\partial_xH(x)\), spatial averaging fixes \(v=\bar{\eta}\), where \(\bar{\eta}=L^{-1}\int_0^L\eta(x)\,dx\).
One integration in the comoving frame then gives
\begin{equation}
  m(x)^{2n-1}=m(0)^{2n-1}+B(x),
  \qquad
  B(x)\equiv\int_0^x[\bar{\eta}-\eta(y)]\,dy .
  \label{eq:alm-brownian-slope}
\end{equation}
Since \(B(0)=B(L)=0\), \(B(x)\) is a Brownian bridge and Eq. \eqref{eq:alm-brownian-slope} implies \(m(L)=m(0)\). We will consider Free-end Neumann boundary conditions, \(m(0)=m(L)=0\).
The physical model therefore corresponds to \(p=(2n-1)^{-1}\) in the generalized stationary process
\begin{equation}
  m(x)=\operatorname{sgn}[B(x)]|B(x)|^p,
  \qquad p>0.
  \label{eq:generalized-brownian-slope}
\end{equation}
We now use the Brownian mapping in Eq.~\eqref{eq:generalized-brownian-slope} to derive the scaling properties of the associated height field,
\(H(x)-H(0)=\int_0^x m(y)\,dy\).

Under \(x\to bx\), the Brownian bridge scales as \(B\to b^{1/2}B\), so \(m\to b^{p/2}m\).
Spatial integration, \(H(x)-H(0)=\int_0^x m(y)\,dy\), then gives
\begin{equation}
  \alpha_m=\frac{p}{2},
  \qquad
  \alpha=1+\frac{p}{2}.
  \label{eq:alm-global-exponents}
\end{equation}

\begin{figure}[t]
  \includegraphics[width=\columnwidth]{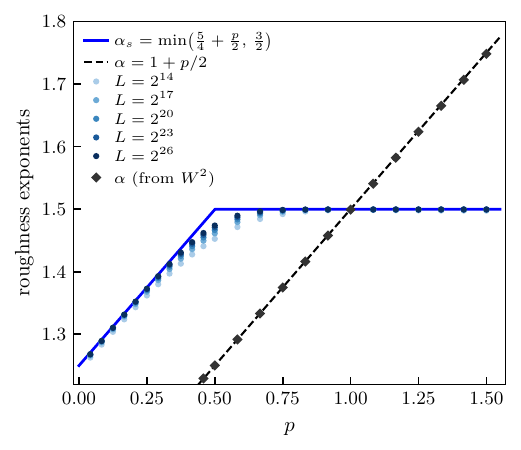}
  \caption{Global and spectral roughness exponents of the generalized stationary slope process as a function of $p$.
  Blue circles show \(\alpha_s\) extracted from the height structure factor, for system sizes \(L=2^{14}\) (lightest) to \(L=2^{26}\) (darkest); the spread near \(p=\frac12\) reflects the logarithmically slow finite-size convergence (End~Matter).
  Black diamonds show \(\alpha\) obtained from the finite-size scaling of \(W^2(L)\) over the same size range.
  The dashed and solid lines show the predictions \(\alpha=1+p/2\) and \(\alpha_s=\min(\frac54+\frac p2,\frac32)\), respectively.
  Free boundary conditions are used; the exponents are independent of this choice (Fig.~\ref{fig:em-bc-test}).}
  \label{fig:alm-exponents}
\end{figure}

To evaluate the two terms in Eq.~\eqref{eq:slope-sector-decomposition}, let
\(\delta_\ell B(x)=B(x+\ell)-B(x)\).
Writing \(B=B(x)\), Eq.~\eqref{eq:generalized-brownian-slope} gives the exact relation
\begin{equation*}
  \Delta_\ell=
  \operatorname{sgn}[B+\delta_\ell B]|B+\delta_\ell B|^p
  -\operatorname{sgn}(B)|B|^p .
\end{equation*}
For \(\ell\ll L\), \(|\delta_\ell B|\sim\ell^{1/2}\), whereas at a typical bulk point \(|B(x)|\sim L^{1/2}\).
Thus typical points satisfy \(|\delta_\ell B|\ll|B|\), and linearizing the exact expression gives
\(\Delta_\ell\simeq p|B|^{p-1}\delta_\ell B\), hence
\(|\Delta_\ell|_{\mathrm{typ}}^2\sim\ell L^{p-1}\).
At points where \(|B(x)|\lesssim|\delta_\ell B|\sim\ell^{1/2}\), however, the linearization fails and the exact expression gives
\(|\Delta_\ell|_{\mathrm r}^2\sim|\delta_\ell B|^{2p}\sim\ell^p\).
These points therefore have a different scaling from typical ones.
For \(p<1\), their increments are sharper by a factor
\(|\Delta_\ell|_{\mathrm r}/|\Delta_\ell|_{\mathrm{typ}}\sim(L/\ell)^{(1-p)/2}\gg1\).
The one-point distribution of \(B(x)\) has width \(L^{1/2}\) and hence a density of order \(L^{-1/2}\) near the origin, so \(P_{\mathrm r}(\ell,L)\sim(\ell/L)^{1/2}\).
These rare points are thus more frequent than the macroscopic jump of columnar KPZ, whose crossing probability is \(\ell/L\).
Substituting the typical and rare estimates into Eq.~\eqref{eq:slope-sector-decomposition} gives
\begin{equation}
  \left\langle|\Delta_\ell|^2\right\rangle
  \sim
  \ell L^{p-1}
  +
  \ell^{p+1/2}L^{-1/2}.
  \label{eq:alm-second-moment}
\end{equation}
The first term is the typical contribution and is always present.
The second is the zero-crossing contribution; for \(p<1\) it corresponds to rare large increments and becomes dominant for \(p<1/2\).
At the marginal value \(p=1/2\), the two contributions have the same algebraic scaling, with a logarithmic correction that causes $\alpha_s$ to converge to $3/2$ anomalously slowly as $1/\ln L$ (End~Matter, ``Finite-size convergence at $p=\frac12$'').
Each term can be matched to the anomalous-scaling form
\(\langle|\Delta_\ell|^2\rangle\sim\ell^{2\alpha_{s,m}}L^{2(\alpha_m-\alpha_{s,m})}\), giving
\(\alpha_{s,m}=p/2+1/4\) when the rare contribution dominates and
\(\alpha_{s,m}=1/2\) when the typical contribution dominates.
Equation~\eqref{eq:height-slope-exponent-relations} then yields all three height exponents in a single expression:
\begin{equation}
  (\alpha,\alpha_s,\alpha_{\mathrm{loc}})
  =
  \begin{cases}
    \left(1+\dfrac p2,\dfrac54+\dfrac p2,1\right),
      & 0<p<\dfrac12,\\[2mm]
    \left(1+\dfrac p2,\dfrac32,1\right),
      & p>\dfrac12.
  \end{cases}
  \label{eq:alm-height-exponents}
\end{equation}
At \(p=1/2\), the two branches meet at  $(\alpha,\alpha_s,\alpha_{\mathrm{loc}})=(5/4,3/2,1)$, with logarithmic corrections to scaling.
For the physical anharmonic Larkin family, \(p\leq1/3\), rare large increments dominate and Eq.~\eqref{eq:alm-height-exponents} gives the exact prediction \(\alpha_s=\alpha+1/4\).
This explains the numerical estimate \(\alpha_s-\alpha\simeq0.23\) previously reported for this model~\cite{Purrello2019Larkin}.
For the numerical comparison, we sample free-BC ground states on both sides of \(p=1/2\), generating the Brownian bridge \(B(x)\) with \(B(0)=B(L)=0\) and setting \(m(x)=\operatorname{sgn}[B(x)]|B(x)|^p\) directly.
Figure~\ref{fig:alm-exponents} compares \(\alpha\) extracted from the finite-size scaling of \(W^2(L)\) and \(\alpha_s\) extracted from the height structure factor \(S_h(k,L)\) with the exact predictions in Eq.~\eqref{eq:alm-height-exponents}.

The visibility of facets correlates with, but is not defined by, which contribution controls the second moment.
For \(p<1/2\), rare large increments dominate and produce a pronounced faceted morphology.
For \(1/2<p<1\), they remain present and dominate sufficiently high moments, but their morphological signature is weaker; the associated bifractality is discussed in the End Matter.
At \(p=1\), typical slope fluctuations recover standard scaling and the interface is super-rough.
For \(p>1\), all positive moments are controlled by typical fluctuations: the interface is nonfaceted and nonbifractal, although its three height exponents remain distinct.

\paragraph{Conclusion.}
We have shown that interfaces with \(\alpha_{\mathrm{loc}}=1\) and \(\alpha_s>1\) admit a natural classification based on the scaling of slope-field increments.
The separation of their global, spectral, and local exponents originates from \(L\)-dependent typical fluctuations, rare large increments whose sampling probability vanishes with system size, or the competition between these contributions.
Columnar KPZ isolates the limit dominated by rare large increments, giving \(\alpha_s=3/2\), whereas the generalized Larkin model displays their coexistence and exchange of dominance for \(0<p<1\), and yields exact exponents across \(p\).
Dominance by rare large increments produces facets, whereas dominance by typical fluctuations need not.
When both scaling sectors coexist, their exchange of dominance with moment order yields bifractal statistics (End~Matter, ``General mechanism''), so \(q\)- and \(L\)-resolved slope increments diagnose the underlying mechanism.

\begin{acknowledgments}
We thank Christophe Texier for bringing the paper by Dumaz to our attention and for discussions on columnar KPZ.
\end{acknowledgments}

\section*{End Matter}

\begin{figure*}[t]
  \centering
  \includegraphics[width=\textwidth]{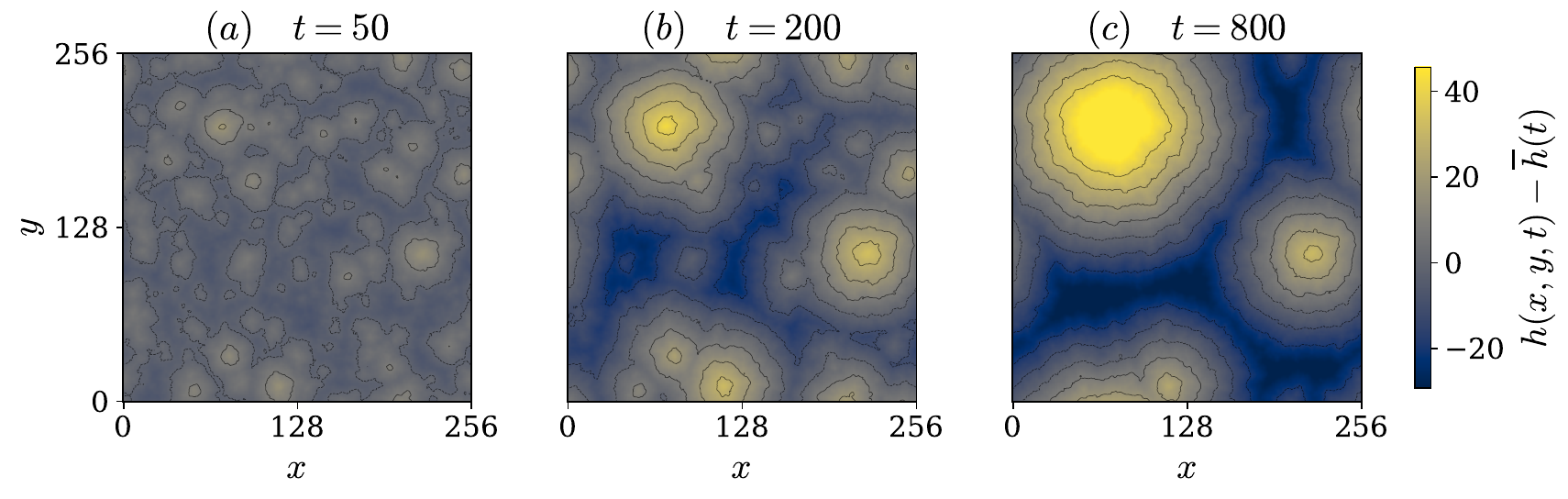}
  \caption{
    Time evolution of a two-dimensional columnar-KPZ interface with columnar disorder for \(L=256\).
    The centered height \(h(\mathbf x,t)-\overline{h}(t)\) is shown at three times.
    Several localized structures compete initially, while a single structure dominates at late times.
  }
  \label{fig:em-kpz2d-localization}
\end{figure*}

\paragraph{Higher dimensions.}
For clarity, consider \(d=2\).
Figure~\ref{fig:em-kpz2d-localization} shows the time evolution of a columnar-KPZ interface for a realization in which the dynamics becomes localized.
Several localized structures compete at early times, producing cone-like profiles, while at late times a single localized structure dominates.

When a finite number of localized structures contributes, neighboring cones meet along lines across which the slope
\(\mathbf m=\boldsymbol{\nabla}h\) jumps.
The total length of these lines scales as \(L\), so a segment of length \(\ell\) crosses one of them with probability
\begin{equation}
  P_{\mathrm r}(\ell,L)
  \sim \frac{\ell L}{L^2}
  \sim \frac{\ell}{L}.
  \label{eq:em-jump-probability-d}
\end{equation}
Consequently, for a displacement \(\boldsymbol{\ell}\) of magnitude \(\ell\),
\begin{equation}
  \left\langle
  |\mathbf m(\mathbf x+\boldsymbol{\ell})-\mathbf m(\mathbf x)|^2
  \right\rangle_{\mathrm r}
  \sim
  \left\langle|\Delta\mathbf m|^2\right\rangle
  \frac{\ell}{L}.
  \label{eq:em-slope-jump-increment-d}
\end{equation}
This is the same \(\ell/L\) dependence as in one dimension and therefore gives \(\alpha_s=3/2\).

More generally, in dimension \(d\), the same result holds whenever the interface contains a finite number of macroscopic facets separated by boundaries of total size \(L^{d-1}\).
Their crossing probability remains proportional to \(\ell/L\), giving \(\alpha_s=3/2\), independently of \(d\).

\paragraph{General mechanism.}\label{para:general-mechanism}
The two-sector decomposition in Eq.~\eqref{eq:slope-sector-decomposition} extends directly to the \(q\)th slope-increment moment:
\begin{equation}
  \left\langle|\Delta_\ell|^q\right\rangle
  \sim |\Delta_\ell|_{\rm typ}^{q}
  +P_{\rm r}(\ell,L)|\Delta_\ell|_{\rm r}^{q}.
  \label{eq:em-moment-decomposition}
\end{equation}
Here \(P_{\rm r}(\ell,L)\) is the sampling probability of rare large increments introduced in Eq.~\eqref{eq:slope-sector-decomposition}; only the characteristic amplitudes of typical fluctuations and rare large increments are raised to the power \(q\).
When each sector is governed by a single characteristic scale, their exchange of dominance with \(q\) produces two scaling branches: low-order moments probe typical fluctuations, whereas sufficiently high-order moments select rare large increments.
The resulting slope statistics are bifractal.

\paragraph{Generalized Larkin model.}
As an exactly solvable example, consider the generalized process in Eq.~\eqref{eq:generalized-brownian-slope} with \(0<p<1\).
Set
\(X=B(x)\), \(Y=B(x+\ell)-B(x)\), and
\(\Delta_\ell=f_p(X+Y)-f_p(X)\), with
\(f_p(b)=\operatorname{sgn}(b)|b|^p\).
At a bulk point, \(|X|\sim L^{1/2}\) and \(|Y|\sim\ell^{1/2}\).
The typical sector follows from linearization,
\(\Delta_\ell\simeq p|X|^{p-1}Y\), and contributes
\begin{equation}
  |\Delta_\ell|_{\rm typ}^{q}
  \sim
  \ell^{q/2}L^{q(p-1)/2}.
  \label{eq:em-typical-moments}
\end{equation}
The linearization fails in the region producing rare large increments, \(|X|\lesssim\ell^{1/2}\), whose probability is \((\ell/L)^{1/2}\).
There \(|\Delta_\ell|\sim\ell^{p/2}\), giving
\begin{equation}
  P_{\rm r}(\ell,L)|\Delta_\ell|_{\rm r}^{q}
  \sim
  \ell^{(1+pq)/2}L^{-1/2}.
  \label{eq:em-rare-moments}
\end{equation}

The two contributions exchange dominance at
\(q_c=1/(1-p)\):
\begin{equation}
  \left\langle|\Delta_\ell|^q\right\rangle^{1/q}
  \sim
  \begin{cases}
    \ell^{1/2}L^{(p-1)/2}, & q<q_c,\\[1mm]
    \ell^{p/2+1/(2q)}L^{-1/(2q)}, & q>q_c.
  \end{cases}
  \label{eq:em-bifractal-moments}
\end{equation}

At \(q=q_c\), a factor \([\log(L/\ell)]^{1/q_c}\) multiplies the common power law.
In particular, for \(q=2\) and \(p=1/2\), this gives the logarithmic correction mentioned in the main text.
The two branches in Eq.~\eqref{eq:em-bifractal-moments} constitute a bifractal spectrum.
For the second moment, \(q=2\), rare large increments dominate for \(p<1/2\), whereas typical fluctuations dominate for \(p>1/2\), reproducing Eq.~\eqref{eq:alm-second-moment}.
At \(p=1\), \(m=B\) and all increments are Gaussian.
For \(p>1\), zero crossings are smoother rather than sharper than typical fluctuations; the second term in Eq.~\eqref{eq:em-moment-decomposition} no longer defines a competing sector of rare large increments, and all positive moments are controlled by the typical contribution.
There is therefore no bifractality.

\begin{figure}[t]
  \centering
  \includegraphics[width=\columnwidth]{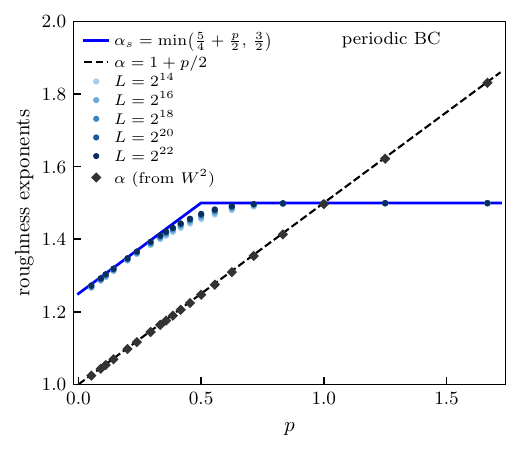}
  \caption{Global ($\alpha$, diamonds) and spectral ($\alpha_s$, circles) roughness exponents versus $p$ for periodic boundary conditions, for system sizes $L=2^{14}$ to $2^{22}$ (light to dark blue). Both exponents agree with the theoretical predictions (lines), confirming that the exponents are insensitive to the choice of boundary conditions.}
  \label{fig:em-bc-test}
\end{figure}

\begin{figure}[t]
  \centering
  \includegraphics[width=\columnwidth]{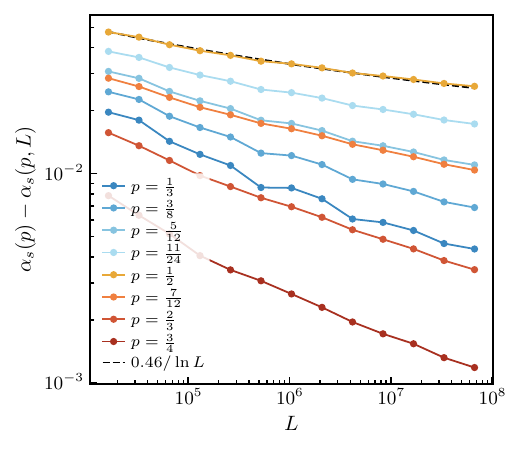}
  \caption{Finite-size convergence of $\alpha_s$ near the crossover $p=\tfrac{1}{2}$, for free boundary conditions and 2000 disorder realizations per $(p,L)$ pair.
  Deviation $\alpha_s(p)-\alpha_s(p,L)$ from the asymptotic prediction versus $L$, for $p$ straddling $\tfrac{1}{2}$ (blue: $p<\tfrac{1}{2}$; amber: $p=\tfrac{1}{2}$; red: $p>\tfrac{1}{2}$).
  The dashed line shows $A/\ln L$: corrections at $p=\tfrac{1}{2}$ track it closely, while corrections decay faster for both $p<\tfrac{1}{2}$ and $p>\tfrac{1}{2}$.}
  \label{fig:em-fsa}
\end{figure}

\paragraph{Second moment and numerical protocol.}
Using the Fourier representation of slope increments, the \(q=2\) results in Eqs.~\eqref{eq:em-typical-moments} and \eqref{eq:em-rare-moments} imply
\begin{equation}
  S_m(k,L)
  \sim
  \begin{cases}
    L^{-1/2}k^{-(p+3/2)}, & 0<p<1/2,\\[1mm]
    L^{p-1}k^{-2}, & p>1/2.
  \end{cases}
  \label{eq:em-slope-structure-factor}
\end{equation}
At \(p=1/2\), the same algebraic \(k^{-2}\) decay is accompanied by the marginal logarithmic correction identified above.
Thus \(\alpha_{s,m}=1/4+p/2\) for \(p<1/2\) and \(\alpha_{s,m}=1/2\) for \(p>1/2\).
Equation~\eqref{eq:height-slope-exponent-relations} then gives \(\alpha_s=1+\alpha_{s,m}\), while spatial integration gives \(\alpha=1+p/2\).
The exponents are insensitive to the choice of boundary conditions, as confirmed in Fig.~\ref{fig:em-bc-test}.
At \(p=\tfrac{1}{2}\), the finite-size corrections to \(\alpha_s\) decay as \(A/\ln L\); Fig.~\ref{fig:em-fsa} shows that corrections decay faster on both sides of the crossover, for \(p<\tfrac{1}{2}\) and \(p>\tfrac{1}{2}\).
The data in Fig.~\ref{fig:alm-exponents} were obtained from independent discrete Brownian bridges for \(L=2^{14},\ldots,2^{26}\), with \(B(0)=B(L)=0\) and \(m(0)=0\), leading to free boundary conditions; the periodic-BC data in Fig.~\ref{fig:em-bc-test} use \(m(0)\) chosen numerically to ensure $\int_0^L m(x)\,dx=0$ which enforces \(h(0)=h(L)\).
We fitted \(\alpha\) from the finite-size scaling of \(W^2(L)\) and \(\alpha_s\) from \(S_h(k,L)\) over the wavevector range \(0.015<k<0.15\).

\bibliography{references}

\clearpage
\onecolumngrid
\setcounter{equation}{0}
\renewcommand{\theequation}{S\arabic{equation}}
\renewcommand{\theHequation}{supplement.\arabic{equation}}

\section*{Supplemental Material}
\label{sec:supplement}

\section{Columnar KPZ with Gaussian columnar disorder}

\subsection{Localized eigenvectors and Gaussian fluctuations}

The low-lying spectrum and localized eigenvectors of the white-noise Anderson Hamiltonian \(\mathcal H\), defined in the main text, were studied in Refs.~\cite{Texier2000EnergyLevels,DumazLabbe2020Localization}.
The result needed here is that a low-lying eigenvector has a dominant localized peak with an exponential envelope and Gaussian fluctuations around it.
Under the Cole--Hopf transformation, the logarithm of the envelope produces the faceted height profile, while the Gaussian fluctuations produce the large-\(k\) regime of its structure factor.
We briefly recall how both features follow from the Riccati representation.
To describe its Dirichlet spectrum, consider for each \(a\) the solution of
\begin{equation}
  \mathcal H f_a=-a f_a,
  \qquad
  f_a(0)=0,
  \qquad
  f_a'(0)=1.
\label{eq:sm-dirichlet-eigenproblem}
\end{equation}
The value \(-a\) is an eigenvalue if and only if \(f_a(L)=0\).
Introduce the Riccati transform \(X_a(x)=f_a'(x)/f_a(x)\).
It starts from \(+\infty\) and satisfies
\begin{equation}
  dX_a=(a-X_a^2)\,dx+g\,dW_x,
  \label{eq:sm-riccati-equation}
\end{equation}
where \(W_x\) is Brownian motion and \(g\) fixes the disorder strength.
Whenever \(f_a\) vanishes, \(X_a\) reaches \(-\infty\) and is immediately restarted at \(+\infty\).
The eigenvalue \(E_n=-a_n\) is selected by requiring the \((n+1)\)-st such explosion to occur at \(x=L\); equivalently, \(\psi_n\) has \(n\) interior zeros.
When \(a=a_n\), \(f_{a_n}\) is proportional to the eigenvector \(\psi_n\).

For \(a>0\), the drift in Eq.~\eqref{eq:sm-riccati-equation} has a stable fixed point \(+\sqrt a\) and an unstable one \(-\sqrt a\).
For fixed \(n\), the low-energy scale satisfies \(a_n\sim a_L\), with \(a_L\sim(\log L)^{2/3}\), and the Riccati process spends most of its time close to the stable fixed point.
A localized peak is created by the rare transition from \(+\sqrt{a_L}\) to \(-\sqrt{a_L}\).
As shown in Ref.~\cite{DumazLabbe2020Localization}, this transition yields
\(X(x)\simeq-\sqrt{a_L}\tanh[\sqrt{a_L}(x-U)]\).
Its inverse Riccati transform gives
\begin{equation}
  \frac{|\psi(U+x)|}{|\psi(U)|}
  \simeq
  \frac{1}{\cosh(\sqrt{a_L}x)}.
  \label{eq:sm-low-eigenvector-peak-shape}
\end{equation}
Thus each peak has exponential tails.

The same fixed-point picture gives the fluctuations around the exponential envelope.
Writing \(X_a=\sqrt a+y\) on a long stable segment and linearizing Eq.~\eqref{eq:sm-riccati-equation} yields
\begin{equation}
  dy=-2\sqrt a\,y\,dx+g\,dW_x,
  \label{eq:sm-riccati-ou}
\end{equation}
an Ornstein--Uhlenbeck process. Consequently,
\begin{equation}
  \ln|\psi(x)|
  \simeq
  -\sqrt a\,|x-U|+\chi(x),
  \qquad
  \chi(x)=\int^x y(s)\,ds.
\label{eq:sm-log-eigenvector-decomposition}
\end{equation}
Since \(y\) is stationary with a finite correlation length, its integral is Gaussian and becomes Brownian on larger scales.
Since the stationary columnar-KPZ profile is proportional to \(\ln|\psi|\), these Brownian-like fluctuations produce the additional \(k^{-2}\) contribution to its structure factor.

\subsection{Finite-time competition and facet coarsening}

At finite time, several eigenvectors contribute to Eq.~\eqref{eq:columnar-kpz-spectral-decomposition}.
Each mode has a dominant peak at its own random center \(U_n\); if \(N\) modes remain relevant, their \(N\) competing exponential envelopes produce a faceted profile, with slope changes where neighboring envelopes exchange dominance.
For any pair \(E_k<E_n\), the relative temporal weight of mode \(n\) with respect to mode \(k\) decays as
\((c_n/c_k)e^{-(E_n-E_k)t}\).
Higher-energy modes and their localization centers therefore disappear progressively, and neighboring facets merge.

To estimate this coarsening, consider two relevant modes \(k<n\), centered at \(U_k\) and \(U_n\) and separated by
\(R=|U_n-U_k|\).
For Gaussian white disorder, one has
\(a_R\sim(\log R)^{2/3}\) and the eigenvector tails decay as \(e^{-\sqrt{a_R}R}\), while the characteristic gap between the competing modes is
\(\Delta E_R\equiv E_n-E_k\sim a_R^{-1/2}\)~\cite{DumazLabbe2020Localization}.
At \(U_n\), the contribution of the lower-energy mode \(k\) is spatially suppressed but gains temporal weight relative to mode \(n\).
Neglecting nonuniversal prefactors and the ratio of initial overlaps \(c_k/c_n\), their ratio scales as
\begin{equation}
  \frac{|c_k\psi_k(U_n)|e^{-E_kt}}
       {|c_n\psi_n(U_n)|e^{-E_nt}}
  \sim
  \exp\!\left[
    -\sqrt{a_R}R+\Delta E_Rt
  \right].
  \label{eq:sm-columnar-mode-competition}
\end{equation}
The center \(U_n\) disappears when the two contributions become comparable.
This occurs after
\begin{equation}
  t_R
  \sim
  \frac{\sqrt{a_R}R}{\Delta E_R}
  \sim
  a_RR
  \sim
  R(\log R)^{2/3}.
  \label{eq:sm-columnar-time-scale}
\end{equation}
Inverting this relation, with \(\log R_t\sim\log t\) at leading order, gives the typical spacing \(R_t\) between the centers that remain relevant at time \(t\), and hence their number:
\begin{equation}
  R_t
  \sim
  \frac{t}{(\log t)^{2/3}},
  \qquad
  N_{\mathrm{rel}}(t,L)
  \sim
  \max\!\left\{
    1,\frac{L(\log t)^{2/3}}{t}
  \right\}.
  \label{eq:sm-columnar-coarsening}
\end{equation}
The number of macroscopic slope changes follows the same law, until only the principal localized mode remains.
To quantify this coarsening, let \(N_{\mathrm{flip}}(t,L)\) be the number of positions at which the slope changes sign.
Since the final stationary profile contains one such sign change, we define
\begin{equation}
  \rho_{\mathrm{flip}}(t,L)
  \equiv
  \frac{\langle N_{\mathrm{flip}}(t,L)\rangle-1}{L}
  \simeq
  \frac{N_{\mathrm{rel}}(t,L)-1}{L}
  \sim
  \frac{(\log t)^{2/3}}{t},
  \qquad N_{\mathrm{rel}}(t,L)\gg1.
  \label{eq:sm-columnar-flip-density}
\end{equation}

\begin{figure}[t]
  \includegraphics[width=\linewidth]{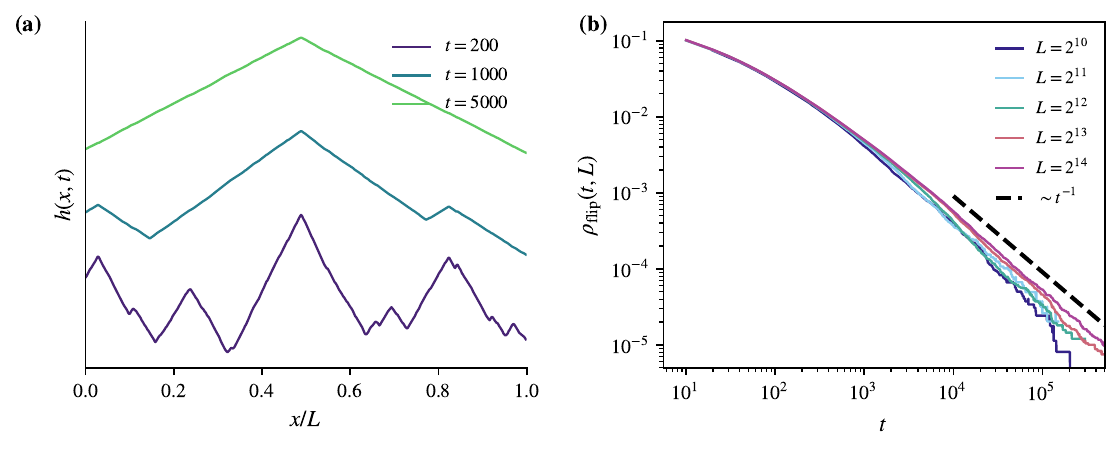}
  \caption{Finite-time coarsening in the one-dimensional columnar KPZ equation.
  (a) Interface profiles at times \(t=200,1000,5000\) for \(L=2^{10}\), vertically translated to display them in the same frame, illustrating the progressive reduction in the number of facets.
  (b) Decay of the density of slope flips for system sizes \(L=2^{10},\ldots,2^{14}\). The dashed line indicates the leading \(t^{-1}\) term in Eq.~\eqref{eq:sm-columnar-coarsening}}
  \label{fig:sm-columnar-kpz-finite-time}
\end{figure}

In the simulations, the slope at site \(i\) is estimated from the height difference between sites \(i-4\) and \(i+4\).
Using points separated by eight lattice spacings prevents small-scale fluctuations from producing spurious sign changes.
Figure~\ref{fig:sm-columnar-kpz-finite-time}(a) shows the time evolution of the interface and the progressive decrease in the number of peaks.
Figure~\ref{fig:sm-columnar-kpz-finite-time}(b) shows the measured density, whose scaling is given by Eq.~\eqref{eq:sm-columnar-flip-density}.

\section{Full slope-increment distribution of the anharmonic Larkin model}

In the End Matter, we obtained the slope multiscaling spectrum by comparing typical intervals with rare intervals containing a small value of the Brownian field.
Here we derive the full distribution of local slope increments and show how the same scaling follows from its tails.
Starting from Eq.~\eqref{eq:alm-brownian-slope}, we write
\[
  m(x)=f_p[B(x)],
  \qquad
  f_p(b)=\operatorname{sgn}(b)|b|^p.
\]
We take \(B\) to be standard Brownian motion.

\subsection*{Calculation of the distribution}

Consider a bulk point \(x\sim L\) and a separation \(1\ll\ell\ll x\), and define
\begin{equation}
  X=B(x),
  \qquad
  Y=B(x+\ell)-B(x),
  \qquad
  \Delta_\ell=f_p(X+Y)-f_p(X).
  \label{eq:sm-XY}
\end{equation}
The variables \(X\) and \(Y\) are independent Gaussians with variances \(x\) and \(\ell\), respectively.
Let
\[
  g_p(u)=f_p^{-1}(u)=\operatorname{sgn}(u)|u|^{1/p}.
\]
For fixed \(X\), the relation
\(\Delta=f_p(X+Y)-f_p(X)\) has the unique solution
\[
  Y=g_p[\Delta+f_p(X)]-X.
\]
Changing variables from \(Y\) to \(\Delta\) therefore gives the exact conditional density
\begin{equation}
  P_\ell(\Delta|X)
  =
  \frac{|\Delta+f_p(X)|^{1/p-1}}{p\sqrt{2\pi\ell}}
  \exp\!\left[
    -\frac{\{g_p[\Delta+f_p(X)]-X\}^2}{2\ell}
  \right].
  \label{eq:sm-exact-conditional-density}
\end{equation}
Averaging over \(X\) yields the full increment distribution,
\begin{equation}
  P_\ell(\Delta;x)
  =
  \int_{-\infty}^{\infty}
  \frac{dX}{\sqrt{2\pi x}}\,
  e^{-X^2/(2x)}
  P_\ell(\Delta|X).
  \label{eq:sm-exact-increment-density}
\end{equation}

For \(r=\ell/x\ll1\), most intervals satisfy \(|Y|\ll|X|\), so that
\begin{equation}
  \Delta_\ell
  \simeq
  f_p'(X)Y
  =
  p|X|^{p-1}Y.
  \label{eq:sm-linearized-increment}
\end{equation}
Conditioned on \(X\), the increment is then Gaussian with variance
\(p^2\ell|X|^{2p-2}\).
Introducing
\[
  \Delta_{\mathrm{typ}}
  =
  p\,\ell^{1/2}x^{(p-1)/2},
  \qquad
  y=\frac{|\Delta|}{\Delta_{\mathrm{typ}}},
\]
and averaging over \(X=x^{1/2}v\) gives
\begin{equation}
  P_\ell(\Delta;x)
  \simeq
  \frac{1}{\Delta_{\mathrm{typ}}}F_p(y),
  \qquad
  F_p(y)
  =
  \frac{1}{\pi}
  \int_0^\infty dv\,v^{1-p}
  \exp\!\left[
    -\frac{v^2}{2}
    -\frac{y^2v^{2(1-p)}}{2}
  \right].
  \label{eq:sm-local-scaling-function}
\end{equation}

For \(\bm{p<1}\) and \(y\gg1\), Eq.~\eqref{eq:sm-local-scaling-function} is dominated by \(v\ll1\), corresponding to intervals close to a zero of \(B\).
Dropping \(e^{-v^2/2}\) and setting
\(z=y^2v^{2(1-p)}/2\) gives
\begin{equation}
  F_p(y)
  \sim
  C_p\,y^{-(2-p)/(1-p)},
  \qquad
  C_p
  =
  \frac{
    2^{(2-p)/[2(1-p)]}
  }{
    2\pi(1-p)
  }
  \Gamma\!\left(\frac{2-p}{2(1-p)}\right).
  \label{eq:sm-algebraic-tail}
\end{equation}
The linear approximation requires \(|X|\gg\ell^{1/2}\), so this algebraic tail ends at
\begin{equation}
  y_{\max}
  \sim
  \left(\frac{x}{\ell}\right)^{(1-p)/2},
  \qquad
  \Delta_{\max}
  \sim
  \ell^{p/2}.
  \label{eq:sm-algebraic-cutoff}
\end{equation}
Thus the distribution consists of a typical core of width
\(\Delta_{\mathrm{typ}}\) and a broad algebraic sector terminated by the zero-crossing scale \(\Delta_{\max}\).
Taking moments of this distribution recovers Eq.~\eqref{eq:em-bifractal-moments}: moments below \(q_c=1/(1-p)\) probe the core, whereas moments above \(q_c\) probe the upper cutoff.

For \(\bm{p>1}\), the large-\(y\) behavior of Eq.~\eqref{eq:sm-local-scaling-function} follows from a saddle at
\[
  v_*=(p-1)^{1/(2p)}y^{1/p}.
\]
Laplace's method gives
\begin{equation}
  F_p(y)
  \sim
  A_p\,y^{-(p-1)/p}
  \exp\!\left[-c_py^{2/p}\right],
  \qquad
  c_p=\frac{p}{2}(p-1)^{1/p-1},
  \label{eq:sm-stretched-exponential-tail}
\end{equation}
where \(A_p\) depends only on \(p\).
The stretched-exponential tail leaves all positive moments controlled by the typical scale \(\Delta_{\mathrm{typ}}\), recovering
\(\zeta_m(q)=1/2\) for every \(q>0\); hence there is no multiscaling for \(p>1\).

\end{document}